\documentclass[aps,prd,amsfonts,nofootinbib,supercriptaddress,10pt]{revtex4}
\usepackage{amssymb,amsmath}
\usepackage{epsfig}

\newcounter{fig}

\newcommand{\beq}{\begin{equation}}
\newcommand{\eeq}{\end{equation}}
\newcommand{\bea}{\begin{eqnarray}}
\newcommand{\eea}{\end{eqnarray}}

\begin{document}

\title{Dark energy-like stars from nonminimally coupled scalar field}

\author{Dubravko Horvat\footnote{dubravko.horvat@fer.hr}, Anja Marunovi\'c\footnote{anja.marunovic@fer.hr}}
\affiliation{University of Zagreb,  Faculty of Electrical
Engineering and Computing, Physics Department, Unska 3, 10000 Zagreb, Croatia }

\begin{abstract} \noindent
We show that even a rather \emph{minimal} extension of the Einstein--Hilbert action
by a nonminimal coupling of the scalar field to the Ricci curvature scalar results in configurations that
resemble more the dark energy stars then the ordinary boson stars. Even though many of those
configurations are endowed by negative principal pressures,
the strong energy condition, as a signal of repulsive gravity,
is not significantly violated in these configurations. When imposing restrictions on matter from energy
conditions we find that the maximally allowed masses are shifted to the lower values due to the violation
of the weak and dominant energy conditions. We also calculate the effective compactness and show that
its maximum value is attained in the region of negative pressures,
and is greater then that in ordinary boson stars. Moreover, we develop a universality
technique which allows to efficiently map small configurations,
that are easily solved by numerical methods, to large astrophysical objects.
\end{abstract}

\maketitle

\section{Introduction}

One of the most peculiar predictions of classical general relativity, al least if matter obeys the strong
energy condition (SEC) are black holes.
(According to the SEC the sum of the energy density and pressures, $\rho+\sum p_i\ge 0$, cannot be negative.)
When quantum effects
are included, black holes lead to a number of thermodynamic paradoxes associated with Hawking radiation
and the implied information loss in black hole spacetimes, thus questioning whether
the final stage of a massive star collapse is a black hole, or perhaps some other
as-yet-not-understood dense object, that stops a further collapse. Sakharov was the first that introduced the concept of nonsingular collapse through
 the equation of state for the cosmological dark energy (for which $p\simeq -\rho$)
 as a super-dense fluid~\cite{Sakharov} and then Gliner assumed  that such a fluid could be the final state of gravitational
 collapse~\cite{Gliner}. Inspired by these ideas Mazur and Mottola investigated
 alternative configurations which led to a solution dubbed gravastar
(\emph{gra}vitational \emph{va}cuum \emph{star})~\cite{MazurMottola}. This anisotropic, highly compact astrophysical object
consists of a de Sitter core and through  vacuum phase transition layer
 matches an exterior Schwarzschild spacetime by avoiding an event horizon formation.
 Although gravastar configurations rest upon a very attractive idea,
all these models are macroscopic in the sense that their foundation rest on studying Einstein's theory in presence of a matter fluid
that obeys some phenomenological equation of state, and as such do not have a proper field theoretic foundation.
Both cases -- when the energy density is distributed on thin-shells~(see e.g. \cite{Visser,Visser2}) or when it continuously varies throughout the star~\cite{DeBenedictis,Horvat:Charged,Horvat:Radial}
-- rely on the so-called \emph{Ans\"atze}--approach. In this approach Einstein's equations are solved in presence of a radially
distributed matter fluid, for which an equation of state or some other relation among
the thermodynamic functions (the energy density, the radial and tangential pressures) is provided. All these models are essentially toy models, and they are important
in the sense that they can be used to provide a better understanding of the main characteristics of black hole mimickers.
But, a complete understanding of these objects will be attained only
if we can provide faithful microscopic (field-theoretic) models for these objects.

Apart from a better understanding at the fundamental level,
field theoretic models of highly compact nonsingular objects,
obtained from a suitable lagrangian of interacting matter fields,
can provide a fundamental explanation for the anisotropy in the principal pressures,
which occurs naturally in the stars made of scalar fields, the so-called boson stars.
Boson stars are nonsingular asymptotically flat
solutions of the Einstein-Klein-Gordon field
equations which govern massive complex scalar fields coupled to gravity.
The extensive research started by Kaup~\cite{Kaup},
who has introduced the notion of the
gravitationally bound state of scalar particles. Soon many papers
considering various versions of scalar field configurations appeared~\cite{Ruffini,Colpi,Friedberg,LeePang,Jetzer,Seidel}.
The growing importance of boson stars resulted in
extensive research  which has been reviewed in~\cite{Liebling,SchunckRev,JetzerRev,LiddleRev}.
When the boson star configurations are considered, one immediately
recognizes that a massive scalar field, even if
self-interacting, cannot produce anisotropy which could support an object
with (asymptotically) de Sitter interior.
Albeit boson stars belong to the realm of very compact objects, it turns out that getting
closer to the main black hole features requires modification of general relativity.
Even though Einstein's theory has passed all observational tests in the weak field limit,
the \emph{true} theory of gravity may differ significantly
in the regime of strong gravitational fields. Moreover, large scale cosmological observations
and conceptual difficulties in quantizing general relativity call as well for its modifications.

In this paper we show that even a rather \emph{minimal} extension of the Einstein--Hilbert action
by a nonminimal coupling of the scalar field to the Ricci curvature scalar results in configurations that
resemble more the dark energy stars then the ordinary boson stars. Even though many of those
configurations are endowed by negative principal pressures,
the strong energy condition, as a signal of repulsive gravity,
is not significantly violated in these configurations.
Yet, the maximum effective compactness is attained in the region of negative pressures,
and is greater then that in ordinary boson stars. This fact supports an idea that the
dark energy stars might present a promising black hole mimicker.
While some attempts have been made to study dark energy stars (see for example Refs.~\cite{DESDymnikova,DESDymnikovaGalaktionov,DESLobo:Stable,DESGhezzi,DESRahaman,DESYazadijev}),
which are loosely speaking in literature taken as objects
that contain a negative pressure somewhere in the interior,
no systematic study has been so far performed of whether and
when boson stars in nonminimal setting violate energy conditions.
In this work we fill that gap.

The paper is organized as follows.
In Sec. II we present the basic Einstein equations
for spherically symmetric configurations of a nonminimally coupled
complex scalar field.
In Sec. III a brief description of ordinary boson star solutions is presented which leads to a situation in which
a necessity of additions mechanism is needed. The nonminimal coupling is
introduced in Sec. IV, where we present solutions with required anisotropic
behaviour of pressures, \emph{i.e.} dark energy-like stars comprising negative pressures.
In that section we also perform analysis of the parameter space for which
the weak and dominant energy conditions are violated.
Moreover, we investigate the effective compactness.
Finally, in Sec. V we discuss our results and give an outlook for future work.

\section{Theory behind -- Equilibrium configurations}

For gravity we take the standard Einstein-Hilbert action:
\beq
S_{EH}=\int d^4 x \sqrt{-g}\frac{R}{16\pi G_N},
\label{E-H action}
\eeq
where $G_N$ is the Newton constant, $R$ is the Ricci scalar, and
$g$ is the determinant of the metric tensor $g_{\mu\nu}$, which is given by
\beq
g_{\mu\nu}=\mbox{diag}(-e^{\nu(r)},e^{\lambda(r)},r^2,r^2\sin^2\theta)
\,.
\label{metric}
\eeq
The space-time metric is static and spherically symmetric as we are
interested only in spherically symmetric equilibrium configurations.
For matter we take an action of a complex scalar field
with a mass $m_\phi$ and a quartic
self-interaction $\lambda_\phi$ coupled nonminimally to gravity:
\beq S_\phi=\int d^4 x
\sqrt{-g}\left(-g^{\mu\nu}\partial_\mu\phi^*\partial_\nu\phi-m_\phi^2
\phi^*\phi-\frac {\lambda_\phi}{2} (\phi^*\phi)^2+\xi R
\phi^*\phi\right), \label{Scalar Action} \eeq
where $\xi$ measures the strength of the coupling between scalar field
$\phi$ and gravity {\it via} the  Ricci scalar $R$ and
$\phi^*$ is the complex conjugate of $\phi$.
It is worth noting here
that in order to produce stable configurations, the scalar field
must be complex. According to the \emph{Derrick's theorem}~\cite{Derrick} regular, static,
nontopological, localized scalar field solutions cannot be created by real scalar fields
(see~\cite{Jetzer Instability} and \emph{e.g.}~\cite{Liebling}).\\
The energy-momentum tensor of a complex scalar field is obtained by varying its action
 with respect to the metric tensor $g^{\mu\nu}$:
\bea
T_{\mu\nu}^\phi & =& 2\delta^\alpha_{(\mu}\delta^\beta_{\nu)}\partial_\alpha \phi^* \partial_\beta \phi-g_{\mu\nu}\left[g^{\alpha\beta} \partial_\alpha \phi^* \partial_\beta \phi
+m_\phi^2 \phi^*\phi+\frac 12 \lambda_\phi (\phi^*\phi)^2\right]\nonumber \\
&-& 2\xi\phi^*\phi G_{\mu\nu}+2\xi\nabla_\mu\nabla_\nu(\phi^*\phi)-2\xi g_{\mu\nu}\Box(\phi^*\phi).
\label{energy-momentum tensor}
\eea
By varying now the full action
\beq
S=S_{EH}+S_\phi.
\label{Full Action}
\eeq
with respect to the metric tensor $g^{\mu\nu}$  we obtain the Einstein equations:
\beq
R_{\mu\nu}-\frac 12 g_{\mu\nu} R=8\pi G_N T_{\mu\nu}.
\label{Einstein Eqs}
\eeq
The
Klein-Gordon equation, the equation of motion for the
scalar field $\phi$ (or $\phi^*$),
is obtained
from Bianchi identities or
by varying~(\ref{Full Action}) with respect to $\phi^*$ (or $\phi$), resulting in:
\beq
\left[\Box-m_\phi^2-\lambda_\phi\phi^*\phi+\xi
R\right]\phi=0.
\label{EOM scalar field}
\eeq
In order to proceed we choose a harmonic time-dependence for the
scalar field
\beq
\phi(r,t)=\phi_0(r)e^{-\imath\omega t},\qquad \phi_0(r) \in \mathbb{R}.
\label{ScaF Ansatz}
\eeq
Even though the scalar field that induces the metric is time-dependent,
the energy momentum tensor created by this field is
time-independent and thus leads to time-independent metric
functions. Hence, condition~(\ref{ScaF Ansatz}) does not contradict the Birkhoff theorem.
Furthermore, the same \emph{Ansatz} for the classical field was also used in
Ref.~\cite{Salati} bearing the name \emph{coherent state}, presumably alluding to their resemblance to quantum coherent states.
\footnote{One should keep in mind however that, a scalar field written as in~(\ref{ScaF Ansatz}), apart from the ground state,
can also represent excited states with higher energy, and a particular combination of these states
can indeed form coherent states. In general, these states contain `coherent' radial oscillations, but do not
in the usual sense constitute quantum coherent states.} Highly excited field configurations were used
to explain flat rotation curves inside galactic halos in Ref.~\cite{Salati2}.
Furthermore, in Ref.~\cite{Multistate BS} it was shown that it
is possible to construct a stable multistate boson
star, with coexisting ground and first excited states. \\
Upon inserting the Ans\"atze~(\ref{ScaF Ansatz}) and~(\ref{metric}) into~(\ref{energy-momentum tensor})
one gets for non-vanishing components of the stress energy tensor:
\bea T_t^t&=&\Big(-m_\phi^2-\omega^2
e^{-\nu}-\frac{\lambda_\phi}{2}\phi_0^2\Big)\phi_0^2-e^{-\lambda}(1+4\xi)\phi_0^{\prime\,
2}-2\xi\phi_0^2 G_t^t-4\xi
e^{-\lambda}\left[\phi_0^{\prime\prime}+\left(\frac{\nu^\prime-\lambda^\prime}{2}+\frac
2 r\right)\phi_0^\prime\right]\phi_0 \nonumber\\
&+&2\xi e^{-\lambda}\nu^\prime\phi_0\phi_0^\prime
\label{Ttt}\\
T_r^r&=& \Big(-m_\phi^2+\omega^2
e^{-\nu}-\frac{\lambda_\phi}{2}\phi_0^2\Big)\phi_0^2+e^{-\lambda}\phi_0^{\prime\,
2}-2\xi\phi_0^2 G_r^r-2\xi e^{-\lambda}\left(\nu^\prime+\frac 4
r\right)\phi_0\phi_0^\prime,
\label{Trr}\\
T_\theta^\theta&=&\Big(-m_\phi^2+\omega^2
e^{-\nu}-\frac{\lambda_\phi}{2}\phi_0^2\Big)\phi_0^2-e^{-\lambda}\phi_0^{\prime\,
2}-2\xi\phi_0^2 G_\theta^\theta-4\xi
e^{-\lambda}\left[\phi_0^{\prime\prime}+\left(\frac{\nu^\prime-\lambda^\prime}{2}+\frac
2 r\right)\phi_0^\prime\right]\phi_0\nonumber\\
&&-4\xi \frac{e^{-\lambda}}{r}\phi_0\phi_0^\prime,
\label{Ttheta}\\
T_\phi^\phi&=&T_\theta^\theta. \label{Tphi}
 \eea
Similarly, the scalar field equation of motion~(\ref{EOM scalar field}) becomes:
\beq
\phi_0^{\prime\prime}
+\left(\frac 2 r+\frac{\nu^{\prime}-\lambda^{\prime}}{2}\right)\phi_0^{\prime}
-e^{\lambda}\Big(m_\phi^2+\lambda_\phi\phi_0^2-\omega^2 e^{-\nu}-\xi R\Big)\phi_0=0.
\label{EOM full}
\eeq
By virtue of~(\ref{EOM full}) it is possible to eliminate the
second derivative of the scalar field $\phi_0^{\prime\prime}$ in
the components of the energy-momentum
tensor~(\ref{Ttt}-\ref{Tphi}), leading to the following form for the first two
Einstein equations ($G_\mu^\nu=8\pi G_N T_\mu^\nu$):
\bea \left[1+2\xi(8\pi G_N)\phi_0^2\right]G_t^t&=&8\pi G_N\Big\{\Big(-m_\phi^2-\omega^2
e^{-\nu}-\frac{\lambda_\phi}{2}\phi_0^2\Big)\phi_0^2-e^{-\lambda}(1+4\xi)\phi_0^{\prime\,
2}\nonumber\\
&&\hskip 1.2cm
-4\xi
\left[m_\phi^2+\lambda_\phi \phi_0^2-\omega^2 e^{-\nu}-\xi R \right]\phi_0^2
+2\xi e^{-\lambda}\nu^\prime\phi_0\phi_0^\prime
\Big\},
\label{Gtt Big}\\
\left[1+2\xi(8\pi G_N)\phi_0^2\right]G_r^r&=&8\pi G_N\Big\{
\Big(-m_\phi^2+\omega^2
e^{-\nu}-\frac{\lambda_\phi}{2}\phi_0^2\Big)\phi_0^2+e^{-\lambda}\phi_0^{\prime\,
2}-2\xi e^{-\lambda}\left(\nu^\prime+\frac 4
r\right)\phi_0\phi_0^\prime \Big\}
\,,\quad
\label{Grr Big}
\eea
where
\bea
G_t^t&=&-e^{-\lambda}\left(\frac{\lambda^\prime}{r}+\frac{e^\lambda}{r^2}-\frac{1}{r^2}\right),\\
G_r^r&=&e^{-\lambda}\left(\frac{\nu^\prime}{r}-\frac{e^\lambda}{r^2}+\frac{1}{r^2}\right).
\eea
There is one more independent equation.
Instead of using the $(\theta\theta)$ Einstein equation (or the equivalent $(\varphi\varphi)$ equation),
it is in fact more convenient to use the trace equation, $G_\mu^\mu=-R=8\pi G_N T_\mu^\mu$, leading to:
\begin{equation}
R = \frac{8\pi
G_N\Big\{2 m_\phi^2 \phi_0^2 +2(1+6\xi)\left[e^{-\lambda}\phi_0^{\prime\,2}+(m_\phi^2-\omega^2e^{-\nu}+\lambda_\phi
\phi_0^2)\right]\phi_0^2\Big\}}  {1+2\xi(1+6\xi)8\pi G_N\phi_0^2}.
\label{Ricci Big}
\end{equation}
It is instructive to add a couple of remarks on this equation. For the case of conformal coupling, $\xi=-1/6$, the only non-vanishing term in
the Ricci curvature scalar is the scalar field mass. Hence in the limit of a vanishing scalar field mass, for which the Ricci scalar is zero,
one obtains a conformal gravity limit, as expected.
Nevertheless, for $\xi\ne -1/6$, as we shall see in the subsequent sections, a variety of configurations is possible.\\
Equations~(\ref{Gtt Big}--\ref{Grr Big})
and~(\ref{Ricci Big}) constitute the central equations in this work.

\subsection{Dimensionless variables}
Before we proceed to solving Eqs.~(\ref{Gtt Big}--\ref{Grr Big}) and~(\ref{Ricci Big}),
for the purpose of numerical studies, it is convenient to
work with dimensionless variables/functions. To this purpose we perform the following rescalings:
\bea
&&\frac{r}{\sqrt{8\pi G_N}}\rightarrow x,\qquad  8\pi G_N\phi_0(r)^2 \rightarrow \sigma(r)^2,\nonumber\\
 8\pi G_N R &\rightarrow&  \tilde{R},\qquad 8\pi G_N m_\phi^2 \rightarrow \tilde{m}_\phi^2,\qquad
8\pi G_N \omega^2 \rightarrow \tilde{\omega}^2.
\label{dimensionless variables}
\eea
Upon these transformations all variables/functions get expressed in terms of reduced Planck units:
\bea
 \bar m_P&=&m_P/\sqrt{8\pi}\,,\quad \mbox{with the Planck mass}\quad m_P= \sqrt{\frac{\hbar c}{G_N}}=1.2209\times 10^{19}~{\rm GeV}/c^2 = 2.17651(13)\times 10^{-8}~{\rm kg},
\nonumber\\
\bar l_P&=&\sqrt{8\pi}\,l_P\,,\quad \mbox{with the Planck length}\quad   l_P= \sqrt{\frac{\hbar G_N}{c^3}}=1.616199(97)\times 10^{-35}~{\rm m}.
\label{Planck units}
\eea
The rescaled (dimensionless) differential equations to be solved are then:
\bea
\lambda^\prime&=&\frac{1-e^\lambda}{x}+x\frac{e^\lambda(\tilde{m}_\phi^2+\tilde{\omega}^2 e^{-\nu}+\frac{\lambda_\phi}{2}\sigma^2)\sigma^2+(1+4\xi)\sigma^{\prime\,2}-2\xi\nu^\prime \sigma\sigma^\prime}{1+2\xi\sigma^2}\nonumber\\
&&+\frac{4x\xi e^{\lambda}(\tilde{m}_\phi^2-\tilde{\omega}^2 e^{-\nu}+\lambda_\phi \sigma^2-\xi\tilde{R})\sigma^2}{1+2\xi\sigma^2},
\label{diff eq lambda}\\
\nu^\prime&=&\frac{(e^\lambda-1)(1+2\xi\sigma^2)/x+x e^\lambda(-\tilde{m}_\phi^2+\tilde{\omega}^2 e^{-\nu}-\frac{\lambda_\phi}{2}\sigma^2)\sigma^2+x\sigma^{\prime\,2}-8\xi\sigma\sigma^\prime}
{1+2\xi\sigma^2+2\xi x \sigma\sigma^\prime},
\label{diff eq nu}\\
\sigma^{\prime\prime}&=&-\left(\frac 2 x
+\frac{\nu^{\prime}-\lambda^{\prime}}{2}\right)\sigma^{\prime}+e^{\lambda}(\tilde{m}_\phi^2+\lambda_\phi\sigma^2-\tilde{\omega}^2
e^{-\nu}-\xi \tilde{R})\sigma,
\label{diff eq sigma}
\eea
with the dimensionless Ricci scalar
\beq
\tilde{R}=\frac{2\tilde{m_\phi}^2\sigma^2+2(1+6\xi)
\left[(\tilde{m}_\phi^2-\tilde{\omega}^2e^{-\nu}+\lambda_\phi\sigma^2)\sigma^2+e^{-\lambda}\sigma^{\prime\,2}\right]}
{1+2\xi(1+6\xi)\sigma^2}
\,,
\label{R:rescaled}
\eeq
where now the {\it primes} denote derivatives with respect to $x$.\\
Equations~(\ref{diff eq lambda}--\ref{diff eq sigma}) yield a unique solution (that depends of course on $\sigma_0$)
when subject to the boundary conditions:
\beq
(1) \quad\lambda(0)=0,\qquad (2)\quad \nu(\infty)=0,\qquad
(3) \quad \sigma(0)=\sigma_0, \qquad
(4) \quad \sigma(\infty)=0.
\label{BC}
\eeq
The first boundary condition ensures that the {\it mass function} $m(r)$ defined in terms of the metric function as
\beq
e^{-\lambda}=1-\frac{2 m(r)}{r}=1-\frac{2\tilde{m}(x)}{x}
\eeq
is zero at $r=0$ (or equivalently at $x=0$). The second boundary condition in~(\ref{BC})
ensures asymptotic flatness at large distances,
\beq
e^{\nu(r)}|_{r\rightarrow \infty}=\left(1-\frac{2m(r)}{r}\right)\Bigl|_{r\rightarrow\infty} \rightarrow \,1.
\eeq
The third and fourth boundary conditions in~(\ref{BC})
are typical for boson stars with a positive scalar mass term ($m_\phi^2 > 0$).
~\footnote{On the other hand, the appropriate boundary conditions in
the case of a negative mass term  ($m_\phi^2 < 0$) are $\sigma(0)=0,\, \sigma(\infty)=\sigma_0$,
resulting topologically stable configurations known as
\emph{global strings}. More generally, a multicomponent scalar field
with appropriate boundary conditions can generate global topological defects
which have been extensively studied in cosmology.}
Equations~(\ref{diff eq lambda}--\ref{diff eq sigma}) together with the boundary conditions~(\ref{BC}) constitute
an eigenvalue problem for $\omega$ -- that is, for each central field value $\sigma_0$ there is an unique $\omega$ that
satisfies the given boundary conditions. The ground state is characterized by zero nodes in the field $\sigma(x)$
(defined as the points $x$ where $\sigma(x)=0$),
while the \emph{n}-th excited state has \emph{n}-nodes in $\sigma(x)$.
In this paper, if not explicitly stated otherwise, boson stars in their ground state will be studied.

 We solve these nonlinear, mutually coupled, differential equations numerically by using the software code COLSYS~\cite{Colsys}.

\subsection{Universality}

In order to solve the problem numerically we need to specify the set of parameters $\{\lambda_\phi,\tilde{m}^2,\tilde{\omega}^2,\xi\}$.
However, for a successful numerical integration these parameters cannot be very different from unity.
On the other hand, in physically interesting situations these parameters may wildly differ from unity. For example,
compact stars have radial size that is measured in kilometers, while numerical solutions give objects whose size
is of the order of the Planck length, $l_P \sim 10^{-38}~{\rm km}$, obviously not very useful.
In order to
overcome this impasse, we observe that the dimensionless equations~(\ref{diff eq lambda}--\ref{R:rescaled}) possess a
`conformal' symmetry.
Indeed, Eqs.~(\ref{diff eq lambda}--\ref{R:rescaled}) are invariant under the following conformal transformations
\beq
x\rightarrow \beta x,\quad \lambda \rightarrow \frac{\lambda}{\beta^2},\quad \tilde{R} \rightarrow \frac{\tilde{R}}{\beta^2},
\quad \tilde{m}^2 \rightarrow \frac{\tilde{m}^2}{\beta^2},\quad \tilde{\omega}^2 \rightarrow \frac{\tilde{\omega}^2}{\beta^2},
\quad \sigma\rightarrow \sigma,\quad \xi\rightarrow \xi.
\eeq
How the mass of the whole boson star changes due to these rescalings can be estimated from
the relation
\beq
M\sim\rho R^3,
\eeq
where $\rho$ is the density which can be approximated by the value
of the potential at a scalar field maximum
\beq
\rho\sim V(\sigma_0)\sim \tilde{m}_\phi^2\sigma_0^2+\lambda_\phi\sigma_0^4.
\eeq
On the other hand, from the virial theorem, according to which star's \emph{gradient energy} $\sim$ \emph{potential energy},
the radius of the star, \emph{i.e.} its core in which most of its energy is contained,
can be estimated from
\bea
(\nabla \phi)^2&\sim& V(\phi), \nonumber\\
\frac{\sigma^2}{\tilde{R}^2} &\sim& \tilde{m}_\phi^2\sigma^2+\lambda_\phi\sigma^4.
\eea
This then implies that the mass of a boson star scales as the radius,
 $\tilde M \propto \tilde R$, leading to
\beq
M\stackrel{\sim}{\rightarrow} \beta M.
\eeq
For example, for a compact object whose radial size
is $R\sim 10~{\rm km} = \beta\, \bar l_P$, we obtain that $\beta$ is of the order of $\beta \sim 10^{38}$.
It then follows that the mass of the scalar field changes from $m_\phi\sim \bar m_P$ to $m_\phi \sim 10^{-38} \bar m_P$ and
the coupling constant from $\lambda_\phi\sim 1$ to $\lambda_\phi\sim 10^{-76}$. In light of Eq.~(\ref{Planck units}), the total mass from $M\sim \bar m_P$ changes to $M\sim 0.2\,M_\odot$, where $M_\odot =2\times 10^{30}~{\rm kg}$
is the solar mass.\\
On the other hand, one can start by setting the scalar field mass $m_\phi$ and estimate the resulting star radius and its total mass.
This allows one to build models that can account for astrophysical objects of vastly different sizes,
namely from dark compact objects~\cite{DCO1,DCO2} to galactic dark matter halos~\cite{Salati,Salati2,Salati3}.

\section{Ordinary boson stars: Case of minimal coupling}

Since the properties of the boson stars with quartic self-interaction are quite extensively studied in Ref.~\cite{Colpi},
here we shall only briefly discuss their main characteristics. Perhaps the most peculiar feature of these configurations is the anisotropy in their principal pressures. Whereby in the (usual) fluid approach to compact stars, anisotropy is treated as a rather dubious and speculative concept (\emph{e.g.} in the physics of neutron stars), it appears as a fairly natural property of boson stars. One can verify this by inspecting Eqs.~(\ref{Trr}-\ref{Ttheta}), which for $\xi=0$ yield:
\beq
\Pi=p_t-p_r=-2 e^{-\lambda}\phi_0^{\prime\,2}.
\label{anisotropy BS}
\eeq
Here we have identified the components of the energy-momentum tensor as
\beq
T_\mu^\nu=\mbox{diag}(-\rho,p_r,p_t,p_t),
\eeq
where $\rho$ is the energy density, $p_r$ the radial pressure and $p_t$ is the tangential pressure ($p_t= p_\theta=p_\phi$).\\
From Eq.~(\ref{anisotropy BS}) we see that anisotropy is strictly a negative function of the radial coordinate. This fact entails that, regardless of the coupling strength,
for minimal coupling, one can only create configurations with $p_r\geq p_t$. In the next section we elaborate more on the consequences of this fact.

\begin{figure}
\centering
\includegraphics[scale=0.7]{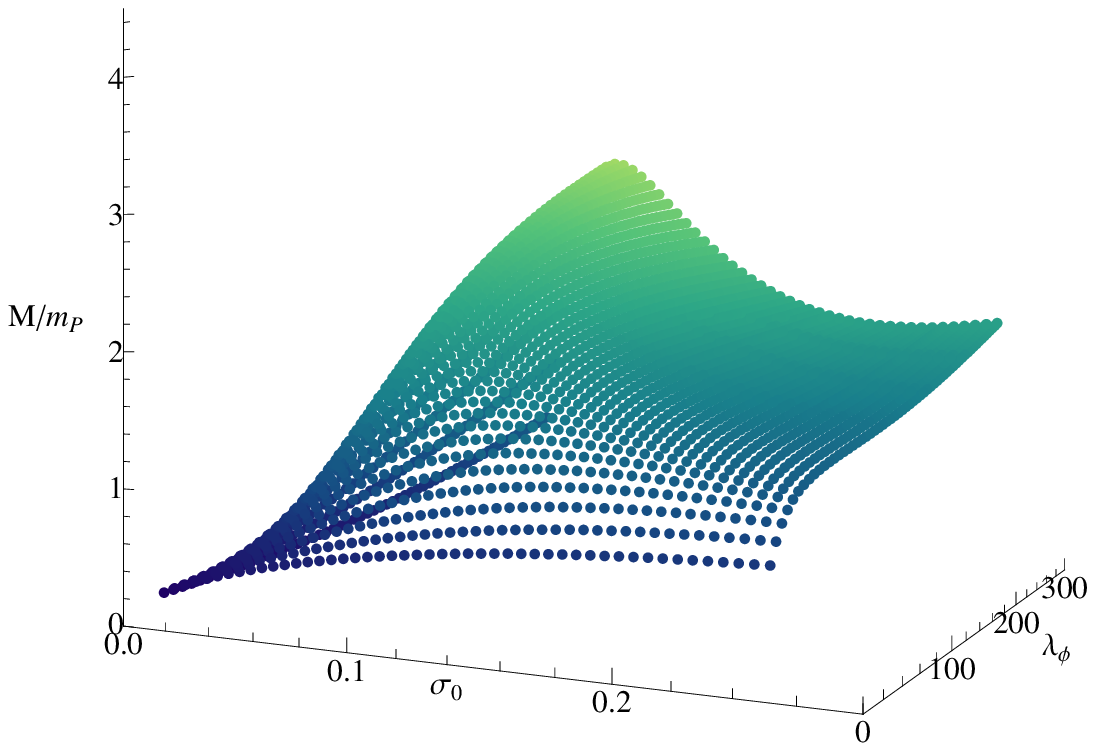}
\includegraphics[scale=0.7]{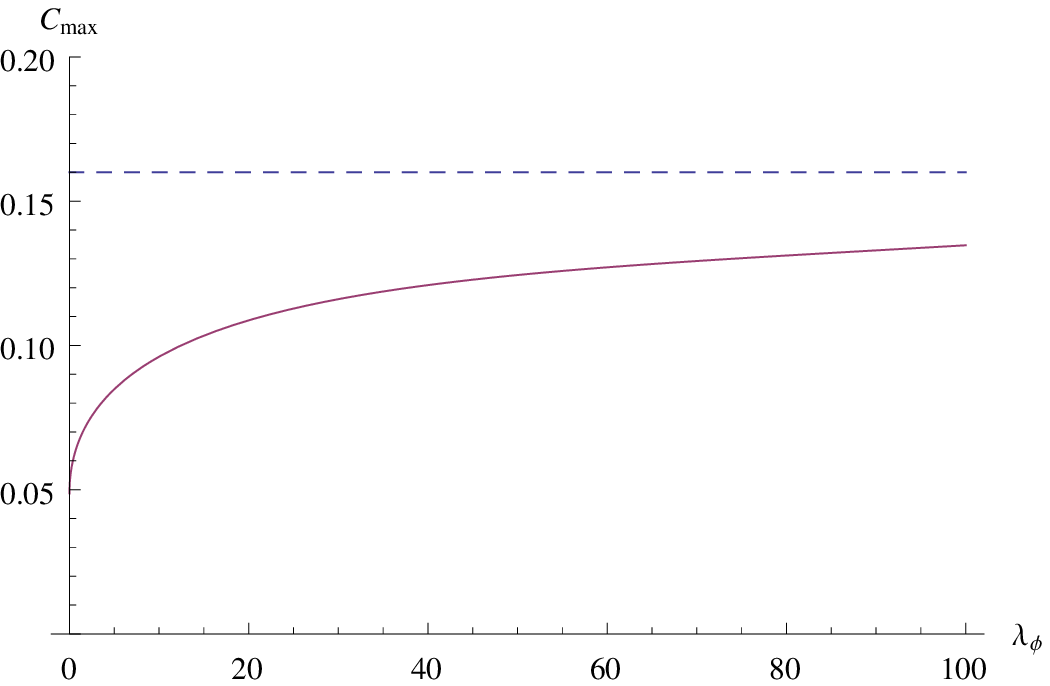}\\
\caption{The star mass as a function of the central field value
$\sigma_0$ and $\lambda_\phi$ for $\xi=0$ on the left panel and the maximal effective compactness as a function of $\lambda_\phi$ on the right panel. Also $m_\phi^2=\bar m_P^2$.}
\label{figMSL3D}
\end{figure}

In order to build a viable astrophysical object, its stability is clearly a basic requirement. Stability of boson stars has been extensively studied in the literature both analytically~\cite{Stability Gleiser,Stability LeePang,Stability Jetzer} and numerically~\cite{d-Stability Choptuik,d-Stability Guzman,d-Stability Seidel}~\footnote{The catastrophe theory is another interesting method that can be found in Ref.~\cite{Stability Schunck}.}. Numerical methods include dynamical evolution of the system at hand, whilst the analytical one rely on the standard Chandrasekhar's methods, \emph{i.e.}
studying the response to a linear perturbation of \emph{static} equilibrium configurations, whereby the total particle number is conserved~\footnote{Since the action~(\ref{Scalar Action}) is invariant under the $U(1)$ symmetry, according to the Noether theorem, there is a conserved (scalar) charge density.}. Both avenues, however, lead to the same conclusion that can be summarized as follows: there exists a critical value of the central field $\sigma_c$ for which the ground state of boson star (nodeless in $\sigma(x)$) will be marginally stable upon small radial perturbations. For this critical field value, the total mass of the star exhibits turnaround in $M(\sigma_0)$-curve (see \emph{e.g.}~\cite{MTW}). Then the configurations left from the peak are stable and those right from the peak are unstable leading to the collapse to a black hole or a dispersion at infinity. An interested reader may find a discussion in Ref.~\cite{Guzman1} on what is the likely fate of the boson star for the right-from-the-peak configurations in the $M(\sigma_0)$-curve. %
In the absence of the self-interaction it was found that the maximally allowed mass is $M_{\rm max}=0.633\, m^2_{\rm Planck}/m_\phi$, and by switching on the self-interaction  it increases as $M_{\rm max}=0.22\sqrt{\Lambda}\,m_{\rm Planck}/m_\phi$, where $\Lambda=\lambda_\phi(4\pi G_N m_\phi^2)$. In the left panel of Fig.~\ref{figMSL3D} we show the star mass as a function of the central field value $\sigma_0$ and $\lambda_\phi$. As the amplitude $\sigma_0$ increases its mass also increases (while its radius decreases). For increasing $\lambda_\phi$ the maximum mass also increases while the critical central field value $\sigma_c$ decreases.

\begin{figure}
\centering
\includegraphics[scale=0.75]{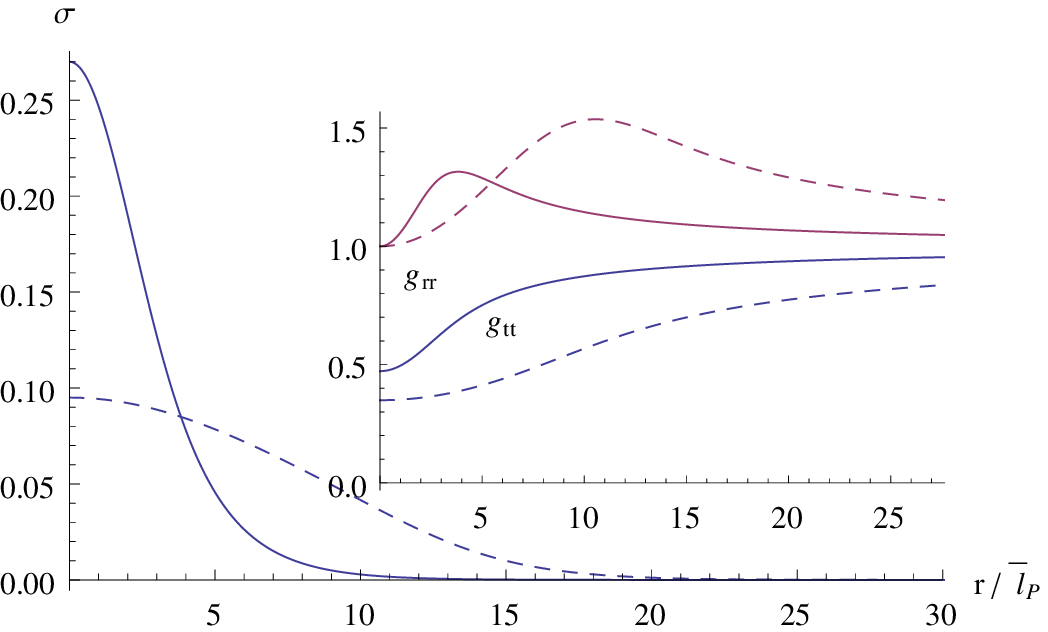}
\includegraphics[scale=0.75]{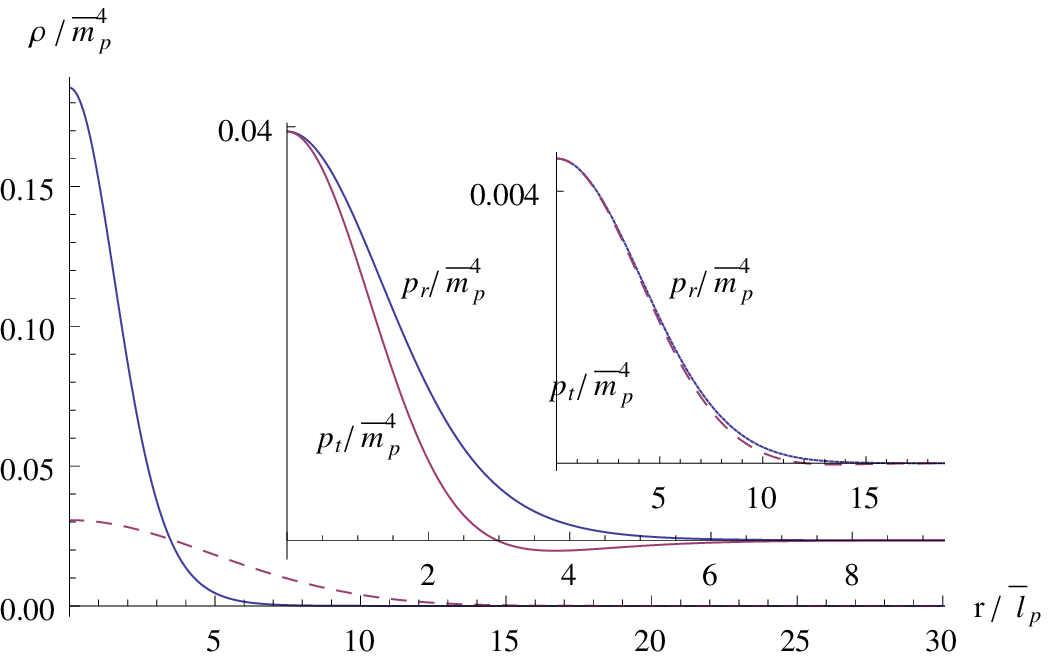}\\
\caption{Left panel: the scalar field as a function of the radial coordinate and in the inset the metric functions $g_{tt}$ and $g_{rr}$.
Right panel: the energy densities  and the principal pressures in the inset. The solid curves are plotted for $\lambda_\phi=0$ ($\{\sigma_c,\, M_{\rm max}\}=\{0.27,\, 0.633\,\bar m_P\}$) and the dashed curves are plotted for $\lambda_\phi=100$ ($\{\sigma_c,\,M_{\rm max}\}= \{0.095,\, 2.257\,\bar m_P\}$). Also $m_\phi^2=\bar m_P^2$ and $\xi=0$.}
\label{Sigmas}
\end{figure}

Albeit self-interacting boson stars have been extensively studied in the literature, the explicit behaviour of their thermodynamic functions, \emph{i.e.} of the energy density and the principal pressures, is lacking. Thereby, here we plot these (and other relevant)
functions for two distinctive regimes: a vanishing self-interaction
and a large self-interaction coupling constant.
For any given value of the coupling constant there is only one configuration that meets those of the maximally allowed mass.
In Fig.~\ref{Sigmas} we show two such configurations for $\lambda_\phi=0$ and $\lambda_\phi=100$. On the left panel,
the profiles of the scalar field and the metric functions (inset) are shown, while on the right panel we show the behaviour of the energy densities
and the corresponding pressures (inset). Two main criteria can be read off from these graphs. The first is the interplay among the central field value and the radius: while one is increasing, the other one is decreasing and {\it vice versa}. An important consequence of this trend is equivalence between the $M(\sigma_0)$ and the $M(R)$-curve. That is, both curves
exhibit turnaround behaviour for equal maximally allowed masses, and hence either can be
used for stability analysis.
Second, the anisotropy in the principal pressures becomes less prominent due to the inclusion of self-interaction.
This behaviour implies that boson stars built from strongly self-interacting fields tend to be more isotropic. Indeed, in this regime, boson stars behave like a polytrope with the equation of state $p_t\approx p_r\propto \rho^{1+1/n}$, where $n$ is the polytropic index.\\
It is also worth noting here that the negative anisotropy~(\ref{anisotropy BS}) cannot rise to more exotic structures with negative principal pressures, found in dark energy stars (\emph{e.g.} gravastars). For the latter, one needs anisotropy to be a positive function of the
radial coordinate (see \emph{e.g.} Refs.~\cite{Cattoen,DeBenedictis}). This is the main reason why we extend this analysis to include nonminimal coupling.

\section{Dark energy-like stars: Effects of nonminimal coupling}

Spherically symmetric static configurations of a nonminimally coupled scalar field modeled by the action~(\ref{Scalar Action}) in the absence of the quartic self-interaction were studied by Bij and Gleiser in Ref.~\cite{Bij Gleiser}. Adopting the $M(\sigma_0)$ stability criterion, the authors calculated the critical (maximally allowed) mass and the critical particle number for a variety of values of the coupling constant $\xi$. The analysis is performed for boson stars both in the ground state (no nodes in the scalar field) as well as in excited states (higher nodes in the scalar field). However, the authors did not analyze the behaviour of the thermodynamic functions, namely of the energy density and pressures.\\
In the case of nonminimal coupling, the anisotropy becomes rather convoluted function of matter \emph{and} geometry
\bea
\Pi&=&-2e^{-\lambda}\phi_0^{\prime 2}-2\xi(G_\theta^\theta-G_r^r)\phi_0^2+2\xi e^{-\lambda}\left(\nu^\prime + \frac 4r\right)\phi_0\phi_0^\prime\nonumber\\
   &-&4\xi\left(m_\phi^2+\lambda_\phi\phi_0^2-\omega^2 e^{-\nu}-\xi R\right)\phi_0^2.
   \label{anisotropy nonmin}
\eea
As mentioned in the previous section, it is likely that Eq.~(\ref{anisotropy nonmin})
may become positive for some radii, which is an important ingredient
of building microscopic configurations with negative pressures.\\
However, when dealing with spherically symmetric, localized, configurations of matter it seems reasonable to invoke the energy conditions as important criteria for physically acceptable
matter.
\subsection{Constraints from the energy conditions}

 Various energy conditions have been proposed as reasonable physical restrictions on matter
fields.
 They originate from the Raychaudhury equation together with the requirement that gravity should be attractive (see \emph{e.g.} Refs.~\cite{Carroll,HorvatEC,DESRahaman}).
 When translated to the energy momentum tensor for an anisotropic matter they read
\bea
\mbox{The Weak Energy Condition (WEC)}&&\quad \rho\ge 0,\quad \rho+p_r\ge 0,\quad \rho+p_t\ge 0,\nonumber\\
\mbox{The Dominant Energy Condition (DEC):}&&\quad \rho-p_r\ge 0,\quad \rho-p_t\ge 0,\nonumber\\
\mbox{The Strong Energy Condition (SEC):}&&\quad \rho+p_r+2 p_t\ge 0.
\label{energy conditions}
\eea
The weak energy condition imposes the requirement of a positive energy density measured by any observer. Also the energy density plus pressures in any direction needs to be positive.
The dominant energy condition requires that the pressures of the fluid do not exceed the energy density, so that the local sound speed in any observable fluid is always less then the speed of light in vacuum. The strong energy condition has very interesting implications. Its violation leads to regions of repulsive gravity such as in cosmological inflation and gravastars.
Hence it is reasonable to require that the WEC and DEC are satisfied by a fluid,
but that the SEC may be violated.

\begin{figure}
\begin{center}
\leavevmode
\includegraphics[scale=0.8]{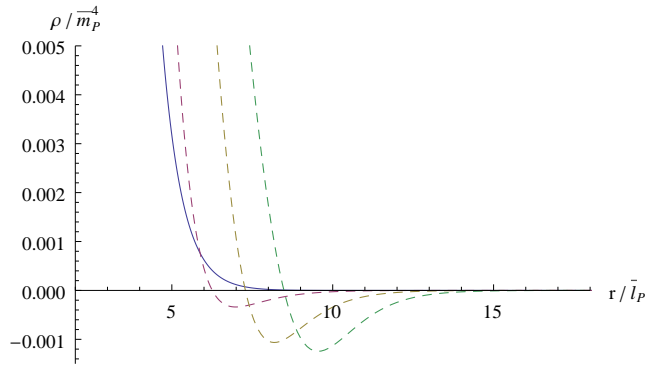}
\end{center}
\caption{The energy density as a function of the radial coordinate. The solid curve stands for $\xi=\xi_{\rm crit}^{\rm WEC}=-1$, the dashed curves are for $\xi=-2, -4, -6$ from left to right, respectively. Also $\lambda_\phi=0$ and $m_\phi^2=\bar m_P^2$.}
\label{figNEC}
\end{figure}

With or without self-interaction it turns out that the WEC is obeyed for
nonminimal couplings only \emph{if} greater than a certain (negative) critical value $\xi>\xi_{\rm crit}^{\rm WEC}$, whereby  $\xi_{\rm crit}^{\rm WEC}$ decreases very slowly as $\lambda_\phi$ increases.
 As an example of the indicated transition, we plot the energy density on Fig~\ref{figNEC} where it is shown that violation of the WEC
 is more prominent as the value of the nonminimal coupling decreases. \\

\begin{figure}
\begin{center}
\leavevmode
\includegraphics[scale=0.8]{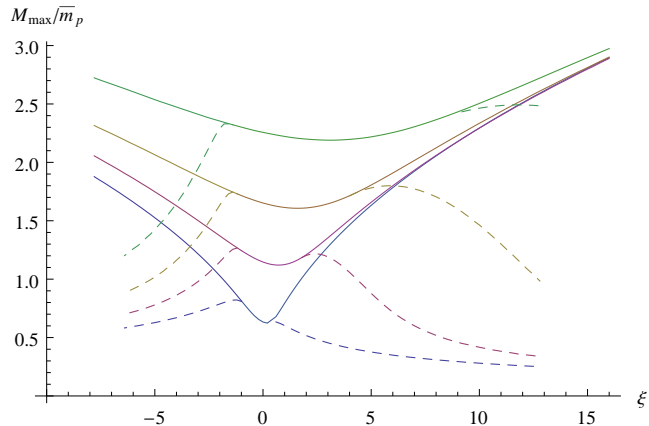}
\end{center}
\caption{The maximum mass as a function of the coupling $\xi$ for
$\lambda_\phi=\{0, 20, 50, 100\}$ from bottom to top. Also $m_\phi^2=\bar m_P^2$. For $\xi<0$ the dashed curves describe configurations that obey the weak energy condition and for $\xi>0$ configurations that obey the dominant energy condition.
The solid curves describe configurations that are not constrained by energy conditions.}
\label{figEC}
\end{figure}

One example of a spacetime that violates the WEC is that of a wormhole (see \emph{e.g.} Refs.~\cite{HoW,Visser3,Visser4,LoboX}). Some other examples would include a more exotic matter.
Although this energy condition is also violated by certain quantum fields, a positive energy density is an essential feature of the classical forms of matter. A consequence of the requirement that the WEC is satisfied is a shift in the "maximally" allowed masses to lower values as depicted on  Fig.~\ref{figEC} for $\xi<\xi_{\rm crit}^{\rm WEC}$.

 In addition, we also require that the energy is not transported faster than light, and hence the dominant energy condition should be satisfied. Another constraint on parameter space emerges from the requirement $\xi>\xi_{\rm crit}^{\rm DEC}$ as shown in Fig.~\ref{figEC}. The dashed curves represent the maximally allowed masses for the by-the-WEC-and-DEC modified configurations, while the solid curves correspond to the old (non-modified) configurations. As such Fig.~\ref{figEC} represents an important result of this paper due to the fact that it establishes new configurations for stars that satisfy the WEC and DEC.

The strong energy condition is also violated for certain nonminimal couplings, $\xi>\xi_{\rm crit}^{\rm SEC}$.  As opposed to the problem of violating the WEC and DEC, a violation of the SEC is actually favorable in building highly compact objects. Namely, a region of a compact object that violates the SEC
exhibits repulsive gravity, which is desirable. Violation of the SEC plays an important role in the early universe cosmology, where it is used to explain
 the origin of Universe's large scale structure generated through matter and gravitational perturbations amplified during a hypothetical inflationary epoch in which the SEC is violated. It is also an essential component
 of gravastars, which in their interior, where $p_r(0)=p_t(0)=-\rho(0)$, strongly violate the SEC. In the case of gravastars,
violation of the SEC is crucial for large values of compactness. Unfortunately, here the SEC is significantly violated only if the DEC is violated. Nevertheless we shall explore some effects of violating the SEC in the next subsection.

\subsection{Energy density and pressures profiles}

 It is now of interest to explore thermodynamic functions, namely the energy density and the principal pressures.\\
 Depending on the strength of the self-interaction, configurations with negative principal pressures emerge, that can be described by the equation of state $p_r\propto -\rho^\beta$. This particular equation of state (EoS) is used to describe \emph{dark energy stars}. Even though these configurations exhibit negative principal pressures, the strong energy condition is not violated thus excluding regions with repulsive gravity.

\begin{figure}
\centering
\includegraphics[scale=0.75]{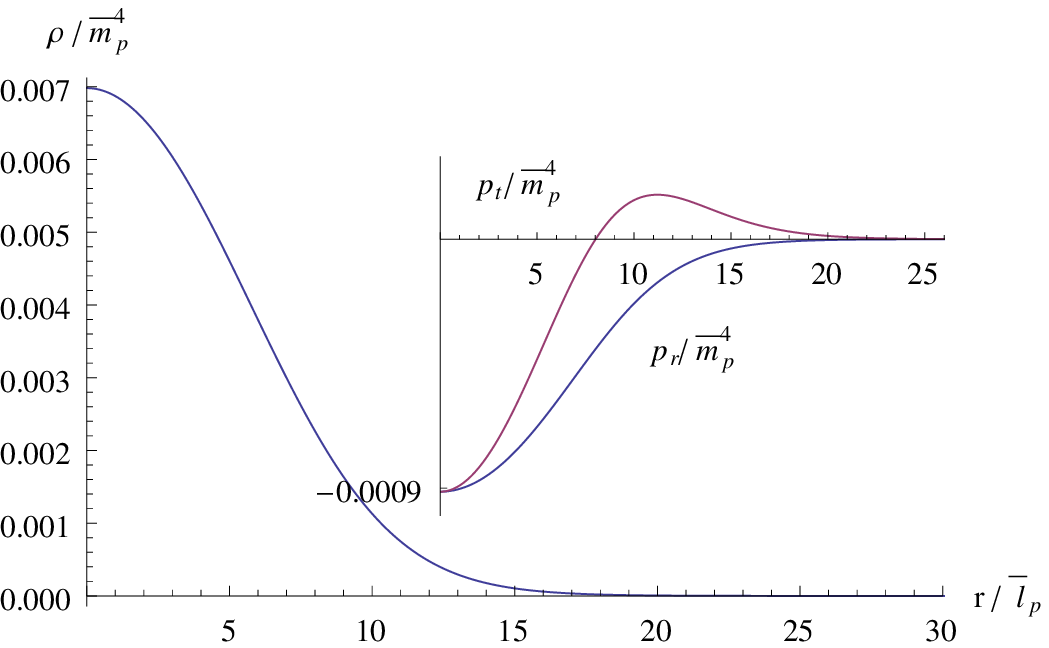}
\includegraphics[scale=0.75]{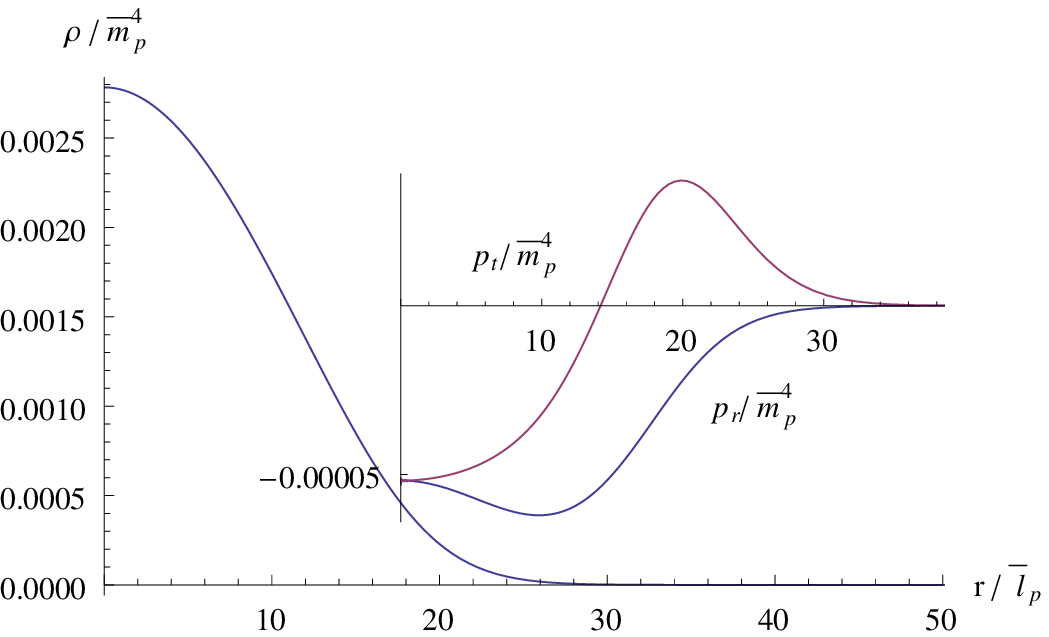}\\
\caption{The energy density and the principal pressures (insets) for $m_\phi^2=\bar m_P^2,\,\xi=-4$
and for
a) $\{\lambda_\phi,\sigma_c\}=\{0,0.050\}$ on the left panel and
b) $\{\lambda_\phi,\sigma_c\}=\{100 ,0.034\}$ on the right panel.}
\label{TDKm}
\end{figure}
\begin{figure}
\begin{center}
\leavevmode
\includegraphics[scale=0.8]{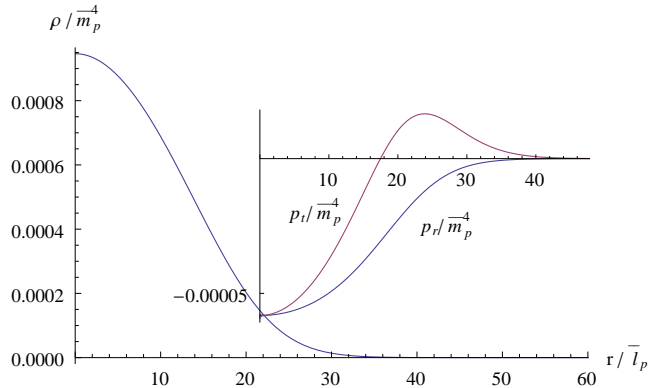}
\end{center}
\caption{The energy density and the principal pressures (inset) for $\xi=-8$, $\lambda_\phi=100$, $\sigma_c=0.02$. Also $m_\phi^2=\bar m_P^2$.}
\label{TDK8L100}
\end{figure}

 One such configuration is shown on the left panel of Fig.~\ref{TDKm}. As a matter of fact, in the absence of self-interaction, for $\xi<\xi_{\rm crit}^{\rm WEC}$ all configurations lying on the $M_{\rm max}(\xi)$-curve can be described by the EoS of a dark energy star, $p_r\propto -\rho^\beta$. When the self-interaction increases, the pressures increase as well, as can be seen on the right panel of Fig.~\ref{TDKm}. Nevertheless, no matter how large the self-interaction is, the dark energy star-like configurations are obtained by choosing an appropriate (\emph{i.e.} negative enough) nonminimal coupling. This effect is clearly shown by comparing Fig.~\ref{TDK8L100} with Fig.~\ref{TDKm}. It is also of interest to observe that the transversal pressures of these configurations, as positive near surface,  are like those of gravastars. This fact brings us to the idea that the gravastars, as not yet formulated within the field theories and as such still of interest to explore,  might be produced in modified gravity that includes higher order terms in the Ricci scalar, Ricci tensor and/or Riemann curvature tensor.

\begin{figure}
\begin{center}
\leavevmode
\includegraphics[scale=0.8]{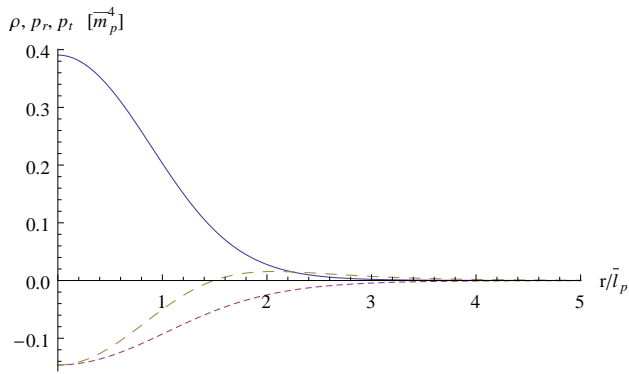}
\end{center}
\caption{The energy density (solid curve), radial (short-dashed curve) and transversal pressure (long-dashed curve) for $\xi=-0.9$, $\lambda_\phi=-30$, $\sigma_0=0.26$. Also $m_\phi^2=\bar m_P^2$.}
\label{figDES}
\end{figure}

However, proper dark energy stars, \emph{i.e.} with negative pressures and violating SEC, can be obtained for $\lambda_\phi<0$. We present one such solution in Fig.~\ref{figDES}. It is interesting that this solution violates only the strong energy condition while
the weak and dominant energy conditions are obeyed. Nevertheless, theories with negative potentials yield Hamiltonians that are unbounded from below, and are at best quasi-stable, \emph{i.e.} field configurations will eventually 'decay' into large fields and roll down to infinity, where energy is minus infinity (see \emph{e.g.}~\cite{Salati2}).

\begin{figure}
\begin{center}
\leavevmode
\includegraphics[scale=0.8]{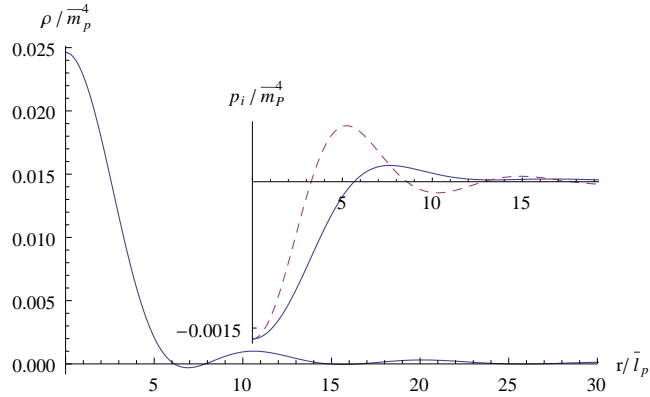}
\end{center}
\caption{The energy density, radial (solid curve in the inset) and transversal pressure (dashed curve in the inset) for $\xi=-0.7$, $\lambda_\phi=0$, $\sigma_0=0.1$ and $m_\phi^2= 0.9\, \bar m_P^2$.}
\label{figGR}
\end{figure}

When excited states of these configurations are considered, the energy density and pressures oscillate in space. Both pressures  are now positive functions of coordinate in the region near the surface thus resembling gravastars solutions. We show one such configuration in Fig.~\ref{figGR}. However, even though stability might not be questionable in this setting, the weak and dominant energy conditions are violated. Yet, it was argued in Ref.~\cite{Salati2} that a galactic halo consisting of highly excited states of ordinary boson stars could explain the rotation of low-luminosity spiral galaxies.

\begin{figure}
\centering
\includegraphics[scale=0.75]{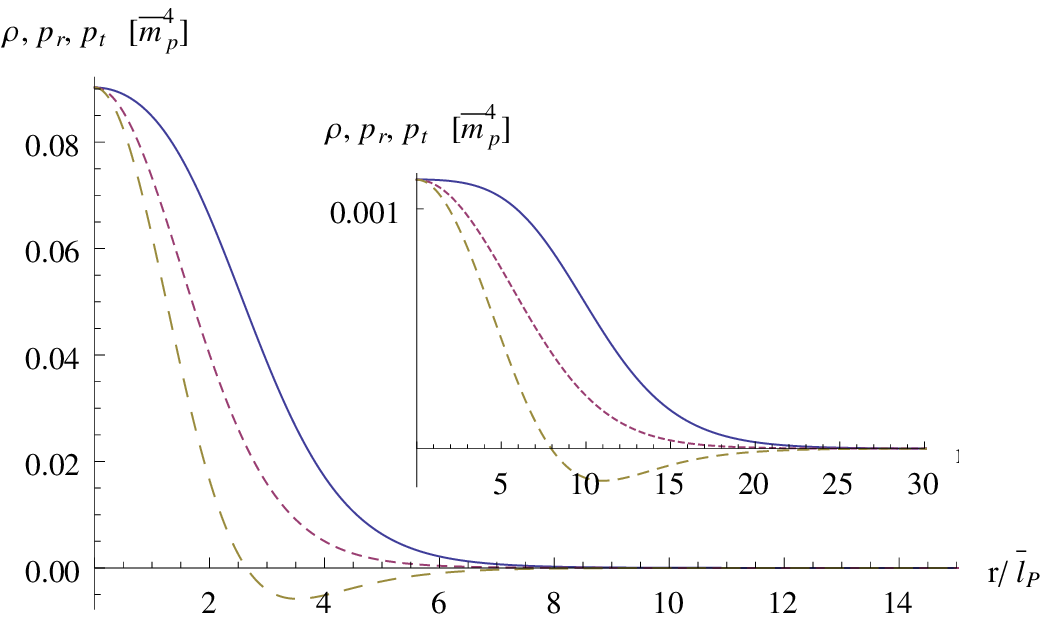}
\includegraphics[scale=0.75]{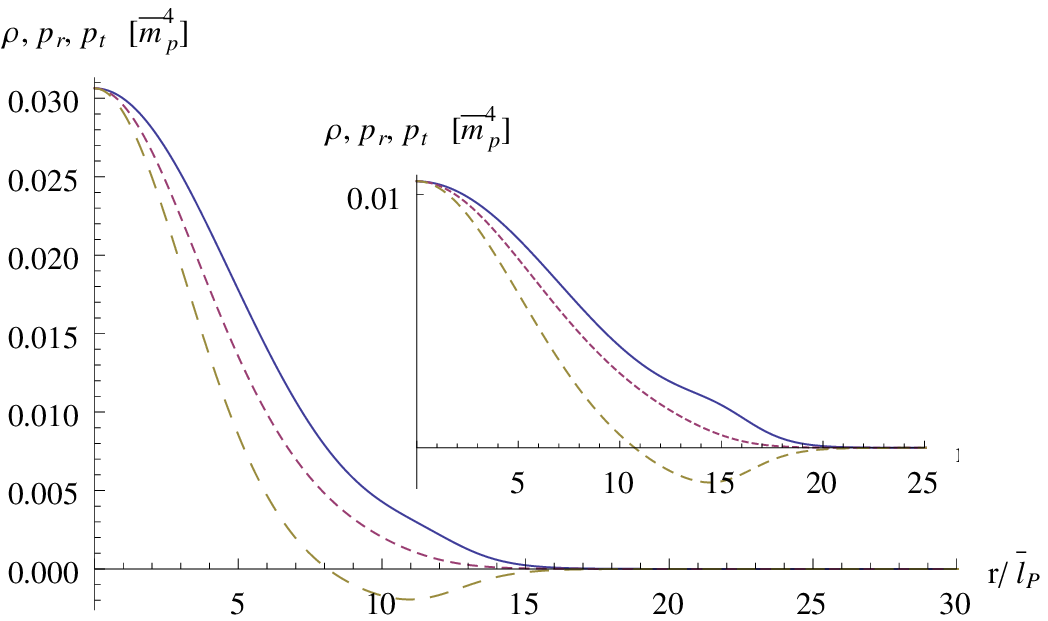}\\
\caption{The energy density (solid), the radial pressure (short-dashed) and the transversal pressure (long-dashed) for a) left panel:
$\lambda_\phi=0$ and $\{\xi,\sigma_c\}=\{0.6,0.2635\}$ and in the inset $\{6.4,0.0364\}$ and b) right panel: $\lambda_\phi=100$ and $\{\xi,\sigma_c\}=\{7.8,0.1194\}$ and in the inset $\{12.8,0.0845\}$. Also $m_\phi^2=\bar m_P^2$. }.
\label{figDec}
\end{figure}

\begin{figure}
\begin{center}
\includegraphics[scale=0.8]{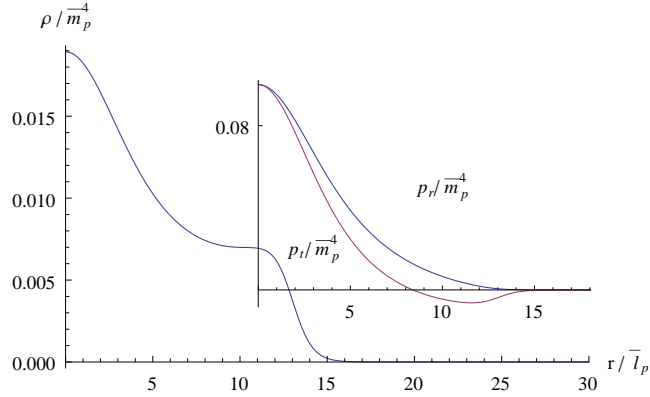}
\end{center}
\caption{The energy density and the principal pressures (insets) for $m_\phi^2=\bar m_P^2,\,\xi=16$, $\{\lambda_\phi,\sigma_c\}=\{0,0.145\}$.}
\label{TDKv}
\end{figure}

Configurations obtained for positive values of nonminimal coupling exhibit positive pressures and hence are quite similar to the ordinary boson stars. In Fig.~\ref{figDec} we plot the energy density, radial and transversal pressure for $\lambda_\phi=0$ on the left panel and $\lambda_\phi=100$ on the right panel. For each $\lambda_\phi$ two configurations are presented, one for $\xi_{\rm crit}^{\rm DEC}$ and the other one for a nonminimal coupling that is much larger then the critical one. For each $\lambda_\phi$ the effect of increasing $\xi$ is only to decrease mass and increase radius (thus decreasing the compactness) without any drastic changes in the behaviour of energy density and pressures. However, the profiles of the energy density and pressures qualitatively do change considerably for different $\lambda_\phi$. On the right panel of Fig.~\ref{figDec} the hump in the energy density occurs as $\xi$ increases. This hump is actually followed by a violation of the strong energy condition which is more significant for larger $\xi$. Hence the hump in the inset of Fig.~\ref{figDec} is more prominent. In order to justify this statement in Fig.~\ref{TDKv} we plot the energy density and pressures for a configuration that strongly violates the SEC (in the region of negative transversal pressure - see inset of Fig.~\ref{TDKv}).

In this subsection, we have seen that static, equilibrium configurations of the self-interacting scalars may exist even though the radial pressure does not exhibit locally positive values. This feature of boson stars is to be contrasted with the stars made of fermionic gas (for which the pressures must be positive and as such help balance matter against gravitational collapse). This ambiguity is clarified if we realize that the (radial) pressure must be positive only if we deal with the fluids, which are obtained as highly excited states of the bosonic or fermionic fields. In the case of the boson stars made of the ground state scalar field configurations, a static, equilibrium solution is obtained only for \emph{charged} fields (i.e. there are no stable (gravitationally bounded) solutions for the real scalar fields), that is, a scalar charge stabilizes the boson star.

Apart from the qualitative behaviour of the energy density and pressures, from this subsection one could also infer subtle relations among the total masses and radii. In particular, increasing the central field value is followed by a decreasing radius up to the maximally allowed mass. This interplay among the mass and the radius is best explored by analysing the effective compactness.
\subsection{Effective compactness}

Following Ref.~\cite{Rezzolla}, we define the \emph{effective compactness} as
\beq
C(\sigma_0,\lambda_\phi)=\frac{M_{99}(\sigma_0,\lambda_\phi)}{R_{99}},
\label{eff. compactness}
\eeq
where $R_{99}$ is the radius at which the mass, defined in terms of the metric function $e^{-\lambda}=1-2m/r$, equals $99 \%$ of the total mass $M=m(\infty)$. The effective radius owes this sort of definition as the scalar field is (exponentially) infinitely extended and thus always with zero compactness.\\
As shown in Ref.~\cite{Rezzolla} the effective compactness in a minimal setting increases with the self-coupling $\lambda_\phi$ and as $\lambda_\phi\rightarrow \infty$ the maximal effective compactness approaches $C_{\rm max}\approx 0.16$ as shown in the right panel of Fig.~\ref{figMSL3D}. For each $\lambda_\phi$, the maximal compactness corresponds to the parameters matching the critical field value $\sigma_c$.
That is the maximally allowed mass and its radius.

\begin{figure}
\begin{center}
\leavevmode
\includegraphics[scale=0.8]{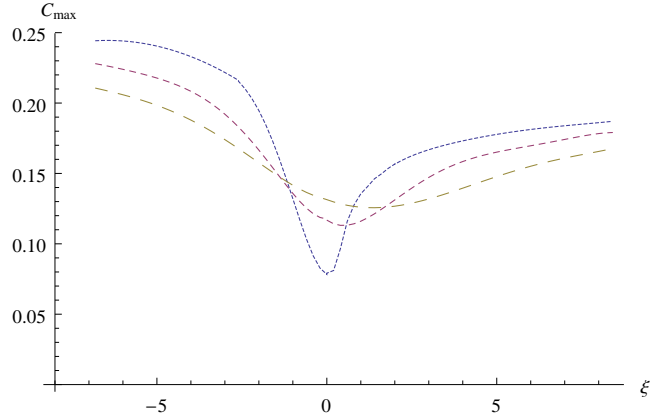}
\end{center}
\caption{The effective compactness as a function of the $\xi$-coupling for $\lambda_\phi=0$ (dotted curve), $\lambda_\phi=20$ (short dashed curve) and $\lambda_\phi=50$ (long dashed curve). Also $m_\phi^2=\bar m_P^2$.}
\label{figCmaxK}
\end{figure}

If $M_{\rm max}$ is not constrained by the weak and dominant energy conditions, the effective compactness for $\xi>0$  is largest for large $\xi$ and $\lambda_\phi=0$ as shown in Fig.~\ref{figCmaxK} and approaches $C_{\rm max}\approx\, 0.20$. This value is only slightly larger then the maximal effective compactness obtained in the minimal setting and can be related to a SEC violation. Why this value is not larger, probably can be explained with the fact that the SEC is violated only near the surface where the transversal pressures become negative.

 However, for $\xi<0$ the compactness is much greater and reaches its maximum value for large negative values of the nonminimal coupling and also in the case when $\lambda_\phi=0$, which approximately equals $C_{\rm max}\gtrsim 0.25$. Even though the strong energy condition is not violated in this region, an increased effective compactness can be attributed to negative pressures that weaken gravity, thus enabling more matter to be accommodated in a fixed volume. This result is also very important as it suggests that, in order to build a highly compact object, we ought to have configurations with negative principal  pressures \emph{and} a violation of the SEC.

\begin{figure}
\begin{center}
\leavevmode
\includegraphics[scale=0.8]{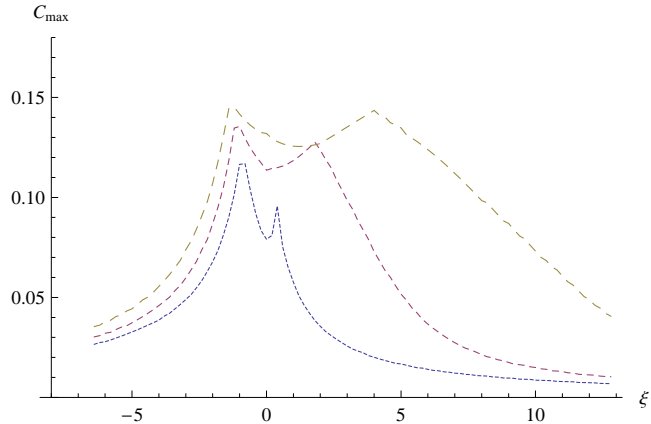}
\end{center}
\caption{The effective compactness as a function of $\xi$ for configurations that obey the WEC and DEC. $\lambda_\phi=0$ (dotted curve), $\lambda_\phi=20$ (short dashed curve) and $\lambda_\phi=50$ (long dashed curve). Also $m_\phi^2=\bar m_P^2$.}
\label{figCmax2}
\end{figure}

When restrictions from the weak and dominant energy conditions are included, the effective compactness behaves as shown in Fig.~\ref{figCmax2}.
The maximal values for each $\lambda_\phi$ are obtained for $\xi_{\rm crit}^{\rm WEC,DEC}$ and then abruptly decrease with incresing/decreasing nonminimal coupling. This brings us to the conclusion that the most compact objects are produced in the domain of negative pressures and large self-couplings, with the maximum effective compactness only slightly larger then the minimal case $C_{\rm max}\gtrsim 0.16$.

Nevertheless, figures~\ref{figCmaxK} and~\ref{figCmax2} are very useful, as one can easily relate the mass to the radius. If we want to create a compact object with, for example, a radius of  $R=15\,{\rm km}$, then from $C_{\rm max}$ one can easily read off its mass. The range of effective compactness $C_{\rm max}=0.05-0.25$ correspond to the masses  $M=(0.5-2.5)\,M_\odot$. \\
 However, in order to obtain scalar's masses and self-couplings, one needs to employ the universality described in Sec. II. By fixing
 the radius to, \emph{e.g.}, $R=15\,{\rm km}$ one can calculate $\beta$ for any configuration with differing radii (in the reduced Planck units).
 Then, the scalar's mass and self-coupling are easily obtained by applying the rescaling conditions $m_\phi^2\rightarrow m_\phi^2/\beta^2$ and $\lambda_\phi\rightarrow \lambda_\phi/\beta^2$. By inspection of all above diagrams depicting the energy density and pressures (in previous subsection) it can be inferred that the radii of all given configurations roughly fall within the range $r=(10-40)\,\bar{l}_P$. Hence, if we want to create a star of $R=15\,{\rm km}$ the corresponding $\beta$s are $\beta=(18.6-4.6)\times 10^{36}$ leading to the scalar's masses $m_\phi=(0.27-1.08)\times 10^{-8}\,{\rm eV}$, which could be in the range of the neutrino masses. To calculate the rescaled self-coupling, let us, for convenience take its starting value $\lambda_\phi=50$. After the rescaling we obtain $\lambda_\phi=(14-0.24)\times 10^{-73}$. But of course,
 if one considers a case when the coupling has reached saturation, one could increase the value of
 the un-rescaled $\lambda_\phi$ arbitrarily, which would then yield more reasonable ({\it i.e.} larger) values of the rescaled $\lambda_\phi$.

%
%

%

\section{Conclusions}

In this paper we have examined spherically symmetric configurations made of a scalar field nonminimally coupled to gravity.
Our results are in perfect agreement with those of Colpi and collaborators (Ref.~\cite{Colpi}) for self-interacting minimally coupled scalars and with those of Bij and Gleiser (Ref.~\cite{Bij Gleiser}) for a non-self-interacting, nonminimally coupled scalar field.

We have shown that already a \emph{minimal}  extension of Einstein's theory to the nonminimal coupling could result in radically different configurations from standard boson stars, \emph{i.e.} dark energy-like stars which are characterized by negative principal pressures. Upon investigating the energy conditions in more detail, it turned out that the strong energy condition, which should be violated in
the interior of dark energy stars, and whose violation signals repulsive gravity, is satisfied in the interior of these configurations. However, we  presented an example of a \emph{proper} dark energy star, \emph{i.e.} with negative pressures and a violating SEC, which is obtained for a negative self-interaction. Even though the configuration presented here does not suffer from violation of the weak and dominant energy condition, configurations with negative potentials, in general, are not that appealing due to their stability issues. We also presented one higher mode solution that led to gravastar-like principal pressures. That is, both principal pressures reveal positive atmosphere (region near surface). But, the strong energy condition is obeyed while the weak and dominant conditions are violated, thus again without spacetime regions with repulsive gravity.

 When imposing restrictions on classical matter by energy conditions we found regions of parameter space for which both the weak and dominant energy conditions are violated. In particular, the weak energy condition is violated for all negative values of the nonminimal coupling,
 if it is less then a critical value $\xi<\xi_{\rm crit}^{\rm WEC}$. The dominant energy condition is violated for all positive values of
 the nonminimal coupling if greater then a critical value $\xi>\xi_{\rm crit}^{\rm DEC}$. The consequences of a violation of the WEC and DEC are encoded in the maximally allowed masses that are now shifted to lower values.
 The strong energy condition is violated in the region of
 a positive nonminimal coupling and is followed by humps in the energy density. Even though violation of the energy conditions does not support the view of classical matter, it would be of interest to explore in more details the imprint of a test particle moving in such a background.

Furthermore, we analyzed the effective compactness for configurations that do or do not satisfy the WEC and DEC, and found that the maximum effective compactness is attained in the regimes of negative pressures for non-self-interacting configurations and equals $C_{\rm max}\gtrsim 0.25$ for
configurations that violate the WEC and $C_{\rm max}\gtrsim 0.16$ for those configurations that obey the WEC and DEC. This result sets limits on the boson star mass. For example, when $R=15\,\rm km$ the maximum mass is $M=(2-2.5)\,M_\odot$ which belongs in the domain of neutron stars. Even though the strong energy condition is not violated, an increased maximum effective compactness could be related to the existence of negative pressures.

In addition, we developed a universality condition based upon which one can calculate scalar's masses and self-couplings for all given configurations. Even though in this paper we focused on parameters that yield compact objects, with the universality condition it is possible to extend this analysis to larger structures that match galactic sizes, such as, for example, dark matter halos. An investigation of observational constraints on the model are underway.

Although theories with a nonminimally coupled scalar field represent a simple and quite benign extension of general relativity, they provide a plethora of different interesting astrophysical structures, ranging from isotropic polytropes to highly anisotropic dark energy-like stars.
Nevertheless, within this model it is, in fact, not possible to create a \emph{highly} compact, nonsingular object whose characteristics are arbitrarily close to those of the Schwarzschild black hole. Yet, from this work one can infer that the real black hole mimicker might be produced in the context of modified theories of gravity.

\section*{Acknowledgements}
A.M. would like to thank Tomislav Prokopec for careful reading of the manuscript and for useful suggestions. This work is partially supported by the Croatian Ministry of Science under the project No. 036-0982930-3144 and also by the
CompStar -- Research Networking Programme of the European Science Foundation.

\end{document}